\documentclass[12pt]{article}
\usepackage[english]{babel}
\usepackage{a4wide}
\usepackage{latexsym}
\usepackage{amsfonts}
\usepackage{graphicx}
\date{}
\begin{document}

\title{\bf {\large{Supersymmetry and Polytopes}}}
\author{\normalsize{Luis J. Boya \footnote{luisjo@unizar.es} \footnote{Contribution
 to the Workshop on Geometry and Physics: Supersymmetry. Bilbao, Spain. May 2008}} \\
\normalsize{Departamento de F\'{\i}sica Te\'{o}rica} \\
\normalsize{Universidad de Zaragoza} \\
\normalsize{E-50009 Zaragoza} \\
\normalsize{SPAIN}}

\maketitle

\begin{abstract}

We make an imaginative comparison between the Minimal
Supersymmetric Standard Model and the $24$-cell polytope in four
dimensions, the Octacube.

\end{abstract}

\section{Introduction}

The Standard Model (SM) of particles and forces has a natural
extension incorporating Supersymmetry, in particular there is the
Minimal Supersymmetric Standard Model (MSSM): this scheme requires
128 boson and 128 fermion states in two diffferent sets, the
ordinary particles and the Susy partners. In this report we shall refer mainly
to the three forces of the microworld, excluding gravitation, the graviton,
etc. However, some references to gravity would be unavoidable, mainly
in relation to extant Supergravity models. No one of the Susy
partners has been seen so far, and we still lack $5$ Higgs bosons
to complete the ordinary set of particles; these are hopefully to
be found soon (at least one Higgs!) in the LHC accelerator,
scheduled to start in late 2008. \\

 The question arises whether there
is any geometry behind this sharing of particles and partners, so
the $2^8$ states would reflect some hidden pattern of symmetry in
Nature. In some proposed models this symmetry is already apparent.
For example, in the old (1978) 11-dimensional SuperGravity theory
of Cremmer, Julia and Scherk \cite{Sugra11}, the supermultiplet
($44, 128, 84$, \ or $44 - 128 + 84$), explicitly

\begin{equation}\label{eq:1}
{\rm {Graviton}} \  h(44) - {\rm {Gravitino}} \  \Psi (128) + \ 3
\ _{\bar {}} \  {\rm {form \ C}} (84),
\end{equation}

\noindent has as underlying geometry the Moufang octonionic plane
$OP^2$, in the following sense \cite{Ram}: as a symmetric space,
it is $OP^2=F_4/B_4 $ where $F_4$ represents the compact form of
the second exceptional simple Lie group, and $B_{4}$ is
represented by the spin group in nine dimensions, ${\rm
{Spin(9)}}$; as the Euler number $\chi(OP^2) = 3$, there are
several triplets of representations of ${\rm {Spin(9)}}$ with
matching dimensions, one of them corresponding to the above
splitting (\ref{eq:1}). We use Cartan\'{}s notation in which
$B_n$ stands for the Spin($2n+1$) group as compact connected
and simply connected representative. For the general philosophy of constructing
these ``Euler multiplets'' and other examples see \cite{Brink}.\\

However, the connection of this $11{\rm {d}}$ supermultiplet with
the particles we see in Nature in our $4$ dimension spacetime is
very remote, to say the least; to begin with, it is a supergravity
multiplet, whereas we aim first to understand the
non-gravitation forces. But the number of required states
$(128=2^{7})$ is not far from the observed helicity states of
known particles (123), and this is the inspiration for the
considerations which follow. \\

A few remarks on the geometry underlying (\ref{eq:1}) are in order.
Pengpan and Ramond \cite{Peg-Ram 98} found several of the triplets referred to above,
but these really come from the SO($16$) group, where $16$ = dim $OP^2$ and
dim Spin($16) = 2^{16/2-1} = 128$,
and it turns out that under the reduction Spin$(9)\subset \ $SO$(16)$, we have $128 - 128 =
(44+84) - 128$, that is, our field content in (\ref{eq:1}). A further remark,
 anticipated in \cite{Sugra11}, is the orthosymplectic group OSp$(1|32)$ as
 invariance supergroup of the action related to (\ref{eq:1}).

\section{The MSSM multiplets}

The simplest supersymmetry is perhaps the pattern $\{(8,8) \equiv
( 8 - 8) \}$ of vector $8_{v}$ and (one of the two) spinor $8_{s}$
representations of $O(8)$, part of triality, corresponding (in the
Ramond-Kostant language of \cite{Ram}) to the sphere $S^{8}=
OP^{1}= {\rm {Spin(9)}}/{\rm {Spin(8)}}$; indeed, by squaring this
doublet $ |8_{v} - 8_{s}|^{2}$ we get the $11{\rm {d}}$ triplet of
above, once we ascend to $11$ dimensions and $M$-Theory (the $8$
dimensions here are just the transverse dimensions of the $10$d
superstring theory, of course). It is remarkable that squaring the
square, say $|8_{v} - 8_{s}|^{4}$, and ascending to $12=(10,2)$
dimensions, it
gives a putative matter content for $F$-Theory \cite{boya}. \\

The particles of the Standard Model are: spin-$1$ ``gaugeons'',
carriers of the three microscopic forces, spin-$1/2$ fermions
(divided into quarks and leptons), and putative spin-$0$ Higgses,
only 3 ``seen'' at the moment in the form of longitudinal degrees
of the carriers of the weak force. Do we see any symmetry pattern
in the masses and groupings of the corresponding helicity states?
Yes, perhaps. We have, for  {\it {gaugeons}}   24 degrees of
freedom:

\begin{equation}\label{eq:2}
 \sharp {\rm {spin\ _{\bar {}} \ 1}} \ {\rm {states}} = 2{\cdot}(\sharp SU(3)+\sharp
SU(2)+\sharp U(1))=2{\cdot}(8+3+1)=24
\end{equation}

Of course, 24 is a dear number to mathematicians (if only for
Dedekind{'}s $\eta$-function or the Leech lattice), and to string
theorist (if only for the bosonic string). This numbering counts
the three massive gaugeons as massless (and hence counting 4, not
1; or 8, not 5 Higgses, later). One can also entertain 27 helicity
states for gaugeons and one minimal Higgs; indeed, 27 is also a
distinguished number (fundamental representation of $E_{6}$, and
some curve intersections in algebraic geometry). \\

Of course, superstrings include gravitation and live in $10=8+2=(9, 1)$
dimensions; so we cite here more reasons why the number $24$ is favoured, besides
the bosonic string. (a): in $R^4$, the maximum number of identical spheres touching each other
(the {\it {kissing} } number) is 24 \cite{Con-Slo}.
(b) The Leech lattice, just mentioned, optimizes the sphere packing in
dimension 24, and links up also with another important mathematical
construct, the Monster group (or largest sporadic finite simple group).
The relation is this: the kissing number for the Leech lattice is 196 560.
Now there are {\it {four}} classes of sporadic (i.e. non-generic)
finite simple groups; besides de `` pariah'' class, there are three
consecutive levels: the Mathieu-group(s) level $M_{24}$, related to the exceptional
automorphism of the symmetric group $S_6$; the Conway-group(s)level, associated
to the automorphism group of the Leech lattice, and the Monster-group level.
The Monster group itself, of order $8\times 10^{53}$ approximately, is
constructed (Lepowski; Borcherds) with vertex operators from
 string theory; it first faithful irreducible representation ({\it {irrep}})
 has dimension 196 883, number tantalizing close to the
  kissing number of the Leech lattice. It presents also the ``Moonshine phenomenon'':
  simple combinations of the dimensions of the {\it {irreps}} give the
  coeficients of a very important modular function, the
  so-called $J(\tau)$ function. See \cite{Gan}. \\

    For {\it {fermions}} we have leptons and quarks in three
generations:
\begin{equation}\label{eq:3}
\sharp {\rm {spin\ _{\bar {}} \ 1/2 \  states}} = {\rm {leptons +
quarks}} = 4{\cdot}(2 \times 3)+3{\cdot}4{\cdot}(2 \times 3)=24+72=96
\end{equation}

\noindent where (3) is for generations, (2) for isospin, $4{\cdot}$ for
massive Dirac states and $3{\cdot}$ for color. Again, this supposes
three sets of massive Dirac neutrinos, what still is to be
confirmed experimentally (as the absence of neutrinoless double
beta decay, for example), which is for the moment uncertain. \\

We shall see below what to do with the big number, $96=24{\cdot}(1+3)$. \\

In the SM it is enough a single complex doublet of Higgses, but we
need {\it {two}} doublets in the MSSM, even in the ``ordinary''
sector, because (among other reasons) up- and down-type quarks
couple differently. Accepting this, we have

\begin{equation}\label{eq:4}
\sharp {\rm  {spin\ _{\bar {}} \ 0 \ states}} = {\rm {Higgses \ in
\ the \ MSSM}} = 2{\cdot}2{\cdot}2=8 \\
\end{equation}

\noindent where two complex doublets make up the 8. Again, 8 is a
special number (if only because the dimension of the last
composition algebras!). \\

Now, in the Susy partners sector, separated from the ordinary one
by the so-called $R-$symmetry, and with Susy broken in order to
prevent unseen mass coincidences between ordinary and Susy
particles, we have to have \\

Spin-1/2 \ gauginos (24), Spin-1/2 Higgsinos (8); Spin-0 squarks
(72) and sleptons (24) \\

So 32 fermions and (72+24 = 96) bosons with $R$-number, to match
the opposites in the ordinary sector. Notice $ 32 = 2^{5}$ appears
in strings again (the gauge group $O(32)$ in heterotic and open
strings); note also the satisfactory feature that the Susy
partners do {\it {not}} introduce new forces, because there are
no more expected spin-$1$ fields.\\

\section{The Polytopes}

From the above counting the reader should keep mainly the numbers
$24$ and $96 =24{\cdot}(1+3)$ in mind. We seek for some {\it {discrete
exceptional}} mathematical objects where these numbers would
appear. Discreteness is obvious, and exceptional because we
subscribe to the philosophy that the mathematical model of Nature
is likely to be exceptional, not generic, as WE are unique(!).\\

{\it {Polytopes}} are generalization of $2$d polygons and $3$d
polyhedra to higher dimensions; they were systematically
investigated first by Schl\"{a}fli around 1850, and are thoroughly
studied in the book of Coxeter {\cite {Cox}}. Starting with the
triangle $T_2$ and the square $H_2$ in the plane $R^2$, there is a
straightforward generalization in arbitrary dimension $n$ to the
{\it {regular generic}} $n$-polytopes lying in $R^n$: hyper-tetrahedron $T_n$, which
is self-dual (=palindromic in the arrangement of vertices, edges,
..., cells), and hypercube $H_n$, with dual hyper-octahedron
$H^{*}_n$. In {\it {even}} dimensions $2n$, their simplices
(vertices $V$, edges $A$, faces $F$,..., cells, etc.) make up a
``supersymmetric'' {\it {alternate}} sum (to zero) because the
Euler number of odd spheres $S^{2n-1}$ is zero; for example

\begin{equation}\label{eq:5}
\ \ {\rm{in}} \ 2d: \ H_2 \ {\rm{is}} \ (4-4); \ \ {\rm{in}}\ 6d:
\ T_6 \ {\rm{is}} \ (7-21+35-35+21-7)
\end{equation}

If one includes ``1'' for the vacuum (with dimension $-1$!) and
another ``1'' for wholeness (the solid body), there is also
``supersymmetry'' for $n$ odd, of course, as $(+2)$ is the Euler
number for even spheres; for instance

\begin{equation}\label{eq:6}
\ \ 3d:H^{*}_3\ (1-6+12-8+1); \ \ \ 5d:T_5 \ (1-6+15-20+15-6+1).
\end{equation}

This counting works because projecting the simplices to the
circumsphere of the polytope tessellates the sphere. \\

Besides these {\it {generic}} regular polytopes there a few {\it
exceptional} regular polytopes: everybody knows of polygons of $p$ sides,
with $p$ any integer $>2$, with ``Susy'' of type $(p-p \ {\rm
{or}} \ 1-p+p-1)$. The Greeks constructed the icosahedron $Y \
(12, 30, 20)$ and -with more effort- its dual dodecahedron $Y^{*}
\ (20, 30, 12)$; and Schl\"{a}fli determined that there are {\it
{only}} a few more exceptions, all in dimension four; see also
{\cite {Lyn}}. Today we
understand all these exceptions as related to the complex numbers
(dimension $2$) and to the quaternion numbers (dim $4$, descending to $3$).
 For example, $ p$-sided regular polygons lying in $R^2$ are related to
the group $SO(2)=U(1)$ being abelian and divisible (injective in
the category of abelian groups); see \cite{Boy2}, \cite{Conw}.\\

For example, one can inscribe a regular polygon in a circle $S^1=U(1)$, with
rotation symmetry $Z_n$; now $U(1)/Z_n\approx U(1)$, as $U(1)$ is
{\it {injective}} in the category of abelian groups (or $Z$-modules),
in the same way that $Z$ is {\it {projective}}; for these elementary notions
of homological algebra consult {\cite {North}}.
For an alternative viewpoint, in which the division (composition)
algebras come first, see {\cite{Conw}}.\\

Clearly the regular polytopes $\Pi_n$ have a center, and two
related groups: a rotation symmetry
 group Rot($\Pi_n)\subset SO(n)$ and an isometry group
 Iso($\Pi_n)\subset O(n)$, where the index $[O:SO]=2$; for example
 Iso$(T_n)=S_{n+1}$, and Rot$(T_{n})= A_{n+1}$, the
alternating subgroup of the symmetric group. Examples for other
polytopes are (where $\prec $ means semidirect product)

\begin{equation}\label{eq:7}
{\rm {Rot}}(H_3)=(Z_2\times Z_2)\prec S_3. \ \ {\rm{Rot}}(Y=Y_3)=A_5. \ \ \sharp Iso(H_n) = 2^n \times n!
\end{equation}

Coxeter explains wonderfully the concept of {\it {truncation}},
which produces some intermediate, quasiregular polytopes by
``cutting corners''; for example, the ordinary cube $H_3$ and the
ordinary octahedron $H^{*}_{3}$ are dual of each other; so
starting e.g. from the cube we get an hybrid, named {\it
{cubeoctahedron}} $H^{*}H_{3}$, with counting $(12, 24, 14)$,
which is a quasiregular polyhedron with $6$ squares and 8
triangles as faces, known from antiquity. The process generalizes
to arbitrary dimensions and polytopes.

\section{The MSSM and the ``Octacube''}

We focus here in the ONLY case in which the cube-octahedron mixing
of above produces a {\it {regular}} polytope, the so-called {\it
24-cell} or (3,4,3) in Schl\"{a}fli ($p,q,r$) notation; its Coxeter diagram is
that of the Lie group $F_{4}$; A. Ocneanu \cite{Ocn} calls it
\underline {Octacube}, living in 4 dimensions, which we shall
write $H^{*}H_{4}$. The reason why is a regular polytope is
related (besides its origin in the quaternions) to the fact that
 the distance from the centre of the hypercube $H_4$ to the
 vertices is the length of the edges, as $\sqrt{1+1+1+1}=2$. The
 $4$-cube $H_4$ is (16, 32, 24, 8) and the 4-octahedron $H^{*}_4$
 is the dual; the hybrid $H^{*}H_4$ becomes \underline{(24, 96,
 96, 24)}: is regular and selfdual! It is the 24-cell, and
 beautiful projections of it to 3 and 2 dimensions are drawn in
 \cite{Cox}; see also \cite{Ocn}.\\

 The automorphism (isometry) group of the 24-cell is the Weyl
 group of the Dynkin diagram for the exceptional Lie group $F_{4}$
 (again!), of order $1152 =384 \cdot 3$, where $384=2^{4}\cdot
 4!=$ order of Automorphism group of $H_4$; notice $\sharp \rm {Rot}(24\ _{\bar {}} \ \rm
 {cell})= 1152/2=576=24^{2}$, where $24= \sharp {\rm {Rot}}(H_{3})$,
 as Rot$( {H}_3)=S_4$; this is a reminder that
 Spin$(4)=$Spin$(3)\times {\rm {Spin}}(3)$, \cite{Boy2}. This enhanced symmetry, the
 factor of 3, in the $H^{*}H_{4}$  with respect to $H_{4}$, does
 {\it {not}} occur in the other cubeoctahedra, and it will be nice if it
 could be related, through triality, to the number of generations
 in particle physics!\\

 Now 96 and 24 are the same numbers as fermions (96) and gaugeons
 (24) in Nature. Even repeated, as required for the R-sector! Is
 this coincidental? Perhaps, but let us play the game:\\

 If ``24'' correspond to the gaugeons and ``96''$=24 + 24\cdot 3$
  to leptons and quarks, what about the Higgses? We venture to
 associate them to the two ``1'' missing in the whole Susy pattern
 of the 24-cell $(1-24+96-96+24-1)$,  {\it {except}} that the ``1''
 must be ``8'': we have 8 Higgs and 8 Higgsinos; perhaps the
 mismatch 1 vs. 8 has something to do with octonions, but granted,
 this is a point we lack understanding. Accepting that
 suggestion, however, the pattern of the 256 expected helicity
 states in the MSSM would look like (with $\pm$ for Bose/Fermi)

\begin{table}[h]
\begin{center}
\begin{tabular}{|c|c|c|c|c|c|}\hline
-1(8) & +24 & -96  & +96  & -24  & +1(8)  \\ \hline
 &  &  &  &  &   \\
\~{H} & g & q + $\ell$ & \~{q} + $\tilde{\ell}$ & \~{g} & H \\
\hline
Higgsinos  & Gaugeons  & Quarks \& Leptons & sQuarks \&
sLeptons &
Gauginos & Higgs \\
Spin 1/2 & Spin 1 & Spin 1/2 & Spin 0 & Spin 1/2 & Spin 0 \\
\hline
\end{tabular}
\caption{Supposed correspondence of the MSSM with the 24-cell.}
\end{center}
\end{table}

That is our correspondence; it goes without saying this pure
 speculative scheme is meant only to stimulate further thoughts
 and works; in particular, it does not help at all to understand
 the pattern of masses we see. The idea is only that the same
 type of ``palindromic'' symmetry of the 24-cell polytope is
 present in the spectrum of elementary particles in the MSSM; we
 are not yet aiming at the deeper reasons for that.\\

 We stress finally the four features we find fairly unique in our
 suggestion: first, the main numbers 24 and 96 are present
 naturally; second, the repeated pattern (twice of each)
 approaches duplication between particles and their Susy partners.
 Third, the 24-cell polytope is absolutely unique, rather than
 only exceptional. Fourth, there is a hint for family
 triplication, as the group $F_4$, with $3$-Torsion, is related
 to octonion triality; for example, in the "Mercedes" Dynkin
 diagram for $O(8)$ triality is manifest; adding to Spin$(8)$ the three
 equivalent representations, we reach $F_4$; \cite{Spain}.\\

    After the first draft of the paper was sent off, we became aware
 of the (now famous) preprint of G. Lisi {\cite {Lisi}}. In fact,
 we subscribe unconsciously to the Pati-Salam `` leptons as the
 fourth color'' philosophy, as Lisi does; he also uses polytopes
 and the $F_4$ group, and hints to a relation between
 triality and generations. He is more ambitious, though, as he considers
 gravitation as well, but does not adhere to supersymmetry.

 \section{Concluding Remarks}

 We are fully aware of the incompleteness of our
 approach. As said, gravitation has been deliberately left over in this essay. Also,
 as we believe octonions should play a role in the ``final
 theory'' (see e.g. \cite{Boy3}), the preliminary and provisional
 aspects of our considerations should be evident: the largest
 exceptional group $E_{8}$ appears conspicuously in theoretical
 constructions (e.g. in the heterotic string by duplicate, as the
 gauge group in $M$-Theory {\cite{Freed}}, etc.); of course, it plays a
 major role in Lisi theory {\cite {Lisi}}. In fact, octonions
 spring from the triality of the Spin(8) group, although this
 phenomenon does not lead to new regular polytopes but to some
 very special lattices (Gosset, 1897). So we feel the two
 omissions (gravitation and octonions, in particular $E_{8}$)
 should go together. As $E_{6}\subset E_{8}$ naturally, is worth to
 recall that $F_4$ is the subgroup of $E_{6}$ fixed by the
 involutory automorphism of it, a kind of complex conjugation.
 In fact, there is a whole chain of groups/subgroups from $E_8$
 to $SU(2)$ which includes triality, duality, automorphisms, etc.
 \cite{King}.\\

 As a final trait of our incompleteness, we mention that the primes
 2 and 3 enter through duality and triality in e.g. $SU(3)$, $F_4$ or
 $E_6$ as center, conjugation or torsion. The prime 5 appears only in $E_8$ as torsion,
  and the possible relevance of this ``5'' for physics escapes totally from us.
  In this context, see {\cite {Boy4}}.\\

 {\bf {Acknowledgements.}} I thank the organizers of the Bilbao
 meeting, especially M. L. Fernandez, for the very stimulating
 workshop. I thank M. Asorey and J.M. Gracia-Bondia (Zaragoza)
 for useful information and for reading the manuscript. In Austin
 I acknowledge conversations with D. Freed and J. Distler.

 \vfill \eject


\begin{thebibliography}{99}
\bibitem{Spain}
It is curious that the Coxeter-Dynkin diagram for $F_{4}$ is not
simply laced, whereas one associates the ``nicest'' Dynkin
diagrams to the simple-laced ADE series; but, as we say in Spain: ``Lo mejor es
enemigo de lo bueno''.
\bibitem{King}
In fact, there is a full chain of groups from $E_8$ to
$A_{1}=SU(2)$: e.g. from $F_4$ we pass to $O(8)$ as $3 \cdot 8 =
28-4 $ and $F_4$ has $2 \cdot 24 +4$ parameters, with 4 the rank.
Now $G_2$ (dim 14) comes from Spin$(8)/ {\rm {Aut}}$, and again dim
$G_{2} = 2 \cdot 6 +2$ with 2 the rank, so $SU(3)$ appears. Finally
$SU(3)/ {\rm {Aut}}=SO(3)= A_1$. I thank A. King (Bath, U.K.) for some of the
above correspondences.
\end{thebibliography}
\end{document}